\documentclass[twoside,11pt,onehalfspacing]{article}

 \usepackage{fullpage}
  \usepackage{authblk}
  \usepackage{natbib}
  \usepackage{graphicx}
  \pdfoutput=1
\usepackage{lipsum}
\usepackage{caption}
\usepackage{hyperref}
\usepackage{subfigure}
\usepackage{amsfonts}
\usepackage{ulem}
\usepackage{xcolor}
\usepackage{multirow}
\usepackage{graphicx}
\usepackage{amsmath}
\usepackage{amssymb}
\usepackage[small,compact]{titlesec}
\usepackage{amsmath}
\usepackage{siunitx} 
\usepackage{booktabs}
\usepackage{float}
\usepackage{accents}
\usepackage{hhline}
\usepackage{float}
\restylefloat{table}
\floatstyle{plaintop}
\restylefloat{table}
\usepackage{tabularx}
\usepackage[section]{placeins}
\newcommand{\bx}{\mathbf{x}}
\newcommand{\bc}{\mathbf{c}}
\newcommand{\bz}{\mathbf{z}}

\newcommand{\EX}{\mathbb{E}}
\newcommand{\pdata}{\bx \sim p_\text{data}}
\newcommand{\pz}{\bz \sim p_\bz}

\begin{document}
\title{Knowledge-based automated planning with three-dimensional generative adversarial networks }

  \renewcommand*{\Authands}{, }
    \author[1]{Aaron Babier}
  \author[1]{ Rafid Mahmood}
  \author[2]{Andrea McNiven}
  \author[3]{Adam Diamant}
  \author[1]{Timothy C. Y. Chan\vspace*{-0.25cm}}
  \affil[1]{Department of Mechanical \& Industrial Engineering, University of Toronto}
  \affil[2]{Radiation Medicine Program, Princess Margaret Cancer Centre}
  \affil[3]{Schulich School of Business, York University\vspace*{-0.5cm}}

  \date{\normalsize{\texttt{\{ababier,rmahmood\}@mie.utoronto.ca}, \texttt{andrea.mcniven@rmp.uhn.ca}, \texttt{adiamant@schulich.yorku.ca}, \texttt{tcychan@mie.utoronto.ca}}}

\maketitle

\begin{abstract}
\textbf{Purpose:} To develop a knowledge-based automated planning pipeline that generates treatment plans using deep neural network architectures for predicting 3D doses.

\textbf{Methods:} Our knowledge-based automated planning (KBAP) pipeline consisted of a generative adversarial network (GAN) to predict dose from a CT image followed by two optimization models to learn objective function weights and generate fluence-based plans, respectively. We investigated three different GAN models. The first two models predicted dose for each axial slice independently. One predicted dose as a red-green-blue (three-channel) color map, while the other predicted a scalar value (one-channel) for dose directly. The third GAN model predicted scalar doses for the full 3D CT image at once, considering correlations between adjacent CT slices. For all models, we also investigated the impact of scaling the GAN predictions before optimization. Each GAN model was trained on 130 previously delivered oropharyngeal treatment plans. Performance was tested on 87 out-of-sample plans, by evaluating using clinical planning criteria and compared to their corresponding clinical plans. KBP prediction quality was assessed using dose-volume histogram (DVH) differences from the corresponding KBAP plans.

\textbf{Results:} The best performing KBAP plans were generated with the 3D GAN, which predicted one channel dose values followed by scaling. These plans satisfied close to 77\% of all clinical criteria, compared to the clinical plans, which satisfied 64\% of all criteria. Additionally, these KBAP plans satisfied the same criteria as the clinical plans 84\% more frequently compared to the 2D GAN model using three channel dose prediction. The 3D GAN predictions were also more similar to the final plan compared to the other approaches, as it better captured the vertical dosimetric relationship between adjacent axial slices. The deliverable plans were also more similar to their predictions than plans generated using any other GAN-based approach and they better captured implicit constraints associated with physical deliverability.

\textbf{Conclusion:} We developed the first knowledge-based automated planning framework using a 3D generative adversarial network for prediction. Our results based on 217 oropharyngeal cancer treatment plans demonstrated superior performance in satisfying clinical criteria and generated more realistic predictions compared to the previous state-of-the-art.

\end{abstract}

\section{Introduction}

The conventional radiation therapy treatment planning process consists of an iterative, back-and-forth procedure between a treatment planner and an oncologist. The duration of a single iteration, compounded by the number of iterations that may take place, means that it can take several days for a treatment plan to be completed~\citep{Das:2009aa}. Automated treatment planning systems have the potential to replace this conventional approach with a more efficient operational paradigm that reduces plan generation lead time~\citep{Purdie:2014aa}. Hospitals that adopt these techniques should also be better equipped to efficiently produce high-quality treatment plans for complicated sites \citep{Ziemer:2017aa} and serve the growing demand for radiation therapy~\citep{Atun:2015aa}.

Knowledge-based automated planning (KBAP) is a two-step approach to automated radiation therapy treatment planning that first predicts a clinically acceptable dose before using optimization to convert the prediction into a deliverable plan~\citep{McIntosh:2017ab,Wu:2017aa,Babier:2018a,Fan:2018aa}. The prediction component of the pipeline, referred to as knowledge-based planning (KBP), typically uses a library of historical treatment plans to learn the dose characteristics of previously delivered plans. It is essential that this prediction model be accurate as the the final plan quality strongly correlates with prediction quality~\citep{Babier:2018a}. 

Many KBP approaches have been introduced in the literature with the objective of either predicting a dose distribution or a dose volume histogram (DVH)~\citep{PCAZhu:2011aa, PCAYuan:2012aa,appenzoller:2012predicting,Yang:2013aa,Shiraishi:2015aa,Shiraishi:2016ab,Younge:2018aa,Babier:2018b,Fan:2018aa}. While the majority of these methods use hand-tailored or low dimensional features for prediction, recent advances in machine learning have spurred the development of KBP methods that predict full dose distributions using automatically generated, high-dimensional features~\citep{McIntosh:2016aa,nguyen2017dose,GANCER,Fan:2018aa}. The most recent work in this space has focused on neural network-based KBP methods, including convolutional neural networks~\citep{nguyen2017dose}, generative adversarial neural networks (GANs) \citep{GANCER}, and recurrent neural networks~\citep{Fan:2018aa}. These models are trained on a library of historical plans and then predict 2D dose distributions for axial slices of a CT image. A 3D dose distribution is formed by concatenation.

In this paper, we develop the first 3D GAN-based KBP method, which takes as input a 3D image and simultaneously predicts the full 3D dose distribution. We embed our prediction model in a KBAP pipeline for oropharynx treatment planning~\citep{Babier:2018b}. Unlike previously developed KBP methods, our approach uses a patient's entire 3D CT image as input and learns to construct spatial features without human intervention. In doing so, it learns to produce the entire 3D dose distribution rather than separate 2D dose distributions for each axial slice. Further, we improve upon a previously developed 2D approach by specializing the GAN to the radiation therapy context: instead of generating an image of the dose, the GAN predicts the dose to be delivered to each voxel directly. We also investigate the effect of scaling the predictions before optimization.

We apply our approach to a dataset of CT images from 217 patients with oropharyngeal cancer that have undergone radiation therapy. Approximately 60\% of these images are used to train our 3D GAN, which is then used to predict the dose distribution for the remaining out-of-sample patients. The predictions are used as input into an optimization model to generate treatment plans~\citep{Babier:2018a}. We demonstrate, over several computational experiments, that our KBAP-specific modifications of (i) predicting the full 3D dose distribution; (ii) representing dose to each voxel as a scalar versus 
a red-green-blue (RGB) color map; and (iii) scaling the predictions before optimization, results in better out-of-sample performance than state-of-the-art benchmarks.

\section{Methods and Materials}
\begin{figure}[tb]
\centering
         \includegraphics[width=0.93\linewidth]{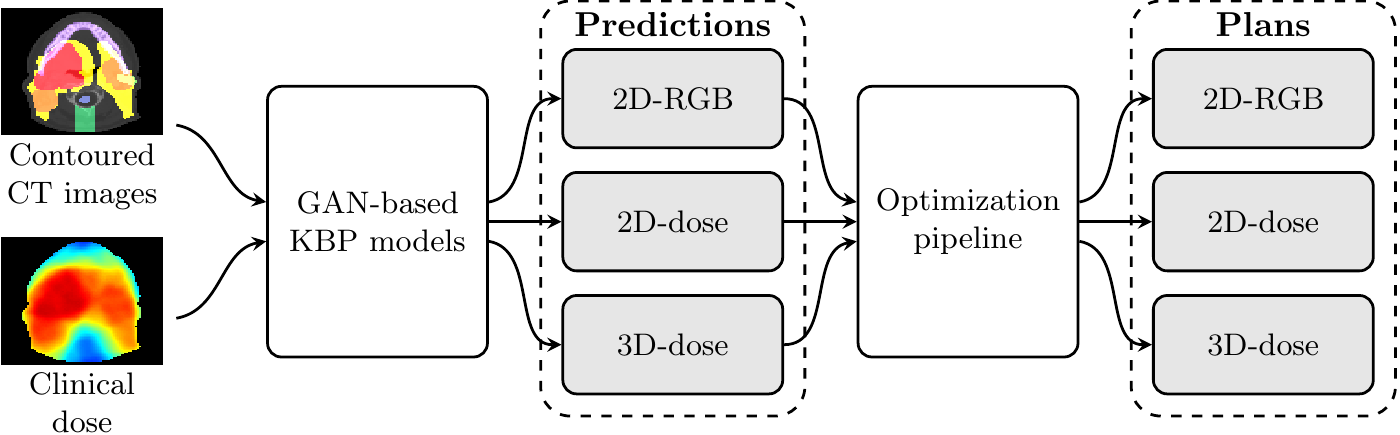}
        \caption{Overview of the knowledge-based automated treatment planning pipeline.}
        \label{fig:GAN}
\end{figure}

We use contoured CT images and dose distributions from clinically accepted treatment plans to train three GAN models in the KBAP pipeline. Each GAN was trained to predict the dose distribution given a contoured CT image. For testing, we passed out-of-sample CT images through each of the GANs to generate dose distributions. These predictions were then passed through an optimization pipeline~\citep{Babier:2018a} to generate the final fluence-based treatment plans. Figure~\ref{fig:GAN} shows a high-level overview of this automated planning pipeline.

\subsection{Prediction Using Generative Adversarial Networks}
A GAN consists of two neural networks known as a \emph{generator} and \emph{discriminator}.~\citep{Goodfellow:2014generative} We focus on \emph{conditional} GANs which, in addition to a Gaussian input, learn to generate different outputs conditioned on known problem-specific characteristics (e.g., CT images)~\citep{isola:2017image}. Specifically, let $\pz$ denote a sample from a Gaussian input. 
The generator network takes as input $\bz$ and a CT image $\bc$ and outputs a predicted dose distribution $\bx = G(\bz, \bc)$. The discriminator network then takes a CT image and the predicted dose distribution as input and outputs a ``belief'' regarding whether the dose distribution is the actual clinical dose (as opposed to artificially produced by the generator). That is, $D(\bx, \bc) \in [0,1]$ where $D(\bx, \bc) = 1$ suggests the discriminator is confident the dose distribution is the clinically delivered dose. Both networks are trained iteratively with a single loss function $\mathcal{L}(D,G)$. Letting $\pdata$ denote the distribution of real delivered plans, we write the training problem as:
\begin{align*}\label{styleTransferLoss}
\min_G \max_D~ \mathcal{L}(D,G)
\end{align*}
where
{\small
\begin{align*}
    \mathcal{L}(D,G) & =\EX_{\pz}\left[ \log (1-D(G(\bz, \bc), \bc)) \right]
    + \EX_{\pdata}\left[ \log D(\bx, \bc) \right] 
    + \lambda \EX_{\pdata,\pz} \left[ \| \bx - G(\bz, \bc) \|_1 \right].
\end{align*}}

The above formulation represents the objective for the most common class of conditional GANs used in problems associated with image generation~\citep{isola:2017image, zhu:2017unpaired}. By minimizing the first term in $\mathcal{L}(D,G)$, the generator learns to construct dose distributions such that $D(G(\bz, \bc), \bc) = 1$. That is, the generator attempts to fool the discriminator into believing that the generated dose is from the real clinical distribution. The discriminator adversarially maximizes the second term in $\mathcal{L}(D,G)$ to output $D(G(\bz, \bc), \bc) = 0$ for $\pz$ and $D(\bx, \bc) =1$ for $\pdata$, i.e., the discriminator attempts to correctly distinguish between artificially generated versus clinically delivered plans. The final term in $\mathcal{L}(D,G)$ is an $l_1$ loss function which forces the generated samples to better resemble the ground truth dataset (i.e., the clinically delivered dose distribution). The hyperparameter $\lambda$ balances the tradeoff between minimizing the GAN loss (first two terms) and having images resemble deliverable plans. 

We constructed generator and discriminator networks derived from the \texttt{pix2pix} architecture proposed in the canonical Style Transfer GAN~\citep{isola:2017image}. The generator possesses a U-net architecture that passes a contoured CT image through consecutive convolution layers, a bottleneck layer, and then several deconvolution layers. The U-net employs skip connections, i.e., the output of each convolution layer is concatenated to the input of a corresponding deconvolution layer. This allows the generator to easily pass ``high-dimensional'' information (e.g., structural outlines) between the inputted CT image and the outputted dose. The discriminator takes as input a dose distribution and CT image and passes them through several consecutive convolution layers until outputting a single scalar value between zero and one. We refer the reader to the original Style Transfer GAN work for full details on the number and size of the convolution and deconvolution layers in the \texttt{pix2pix} architecture~\citep{isola:2017image}.

We trained three GAN variants each with slightly modified architectures; we refer to them as 2D-RGB, 2D-dose, and 3D-dose. The ``2D'' or ``3D'' designation refers to whether dose is being predicted for 2D slices individually or the full 3D image at once. The ``-RGB'' or ``-dose'' designation refers to the input/output channels of the GAN, i.e., whether the output of the generator is a 3-dimensional vector of numbers representing the red-green-blue scale, or a 1-dimensional scalar that represents dose directly (i.e., grayscale). The 2D-RGB GAN is a direct replica of the original \texttt{pix2pix} network and has been previously implemented and tested for KBAP~\citep{GANCER}. It takes as input a single axial slice of the contoured CT image, i.e., a $128 \times 128$ pixel image with three channels for color, and outputs a three channel $128 \times 128$ pixel image that represents dose. Two-dimensional convolution and deconvolution filters are used for the generator and discriminator. The 2D-dose GAN has an almost identical architecture except the output of the generator is a one channel $128 \times 128$ image. In both 2D GANs, a 3D dose distribution is obtained by concatenating all outputted axial slices for a single patient. In contrast, the 3D-dose GAN takes an entire contoured CT image ($128 \times 128 \times 128$ voxels) and generate the corresponding 3D dose distribution. Thus, it uses three-dimensional convolution and deconvolution filters~\citep{hermoza:20173d} with a one channel output to represent dose. Apart from filter dimensions, the 2D-dose and 3D-dose GANs have the same architecture including the same number of layers and filter sizes. 

\begin{figure}[t!]
\centering
        \includegraphics[width=1\linewidth]{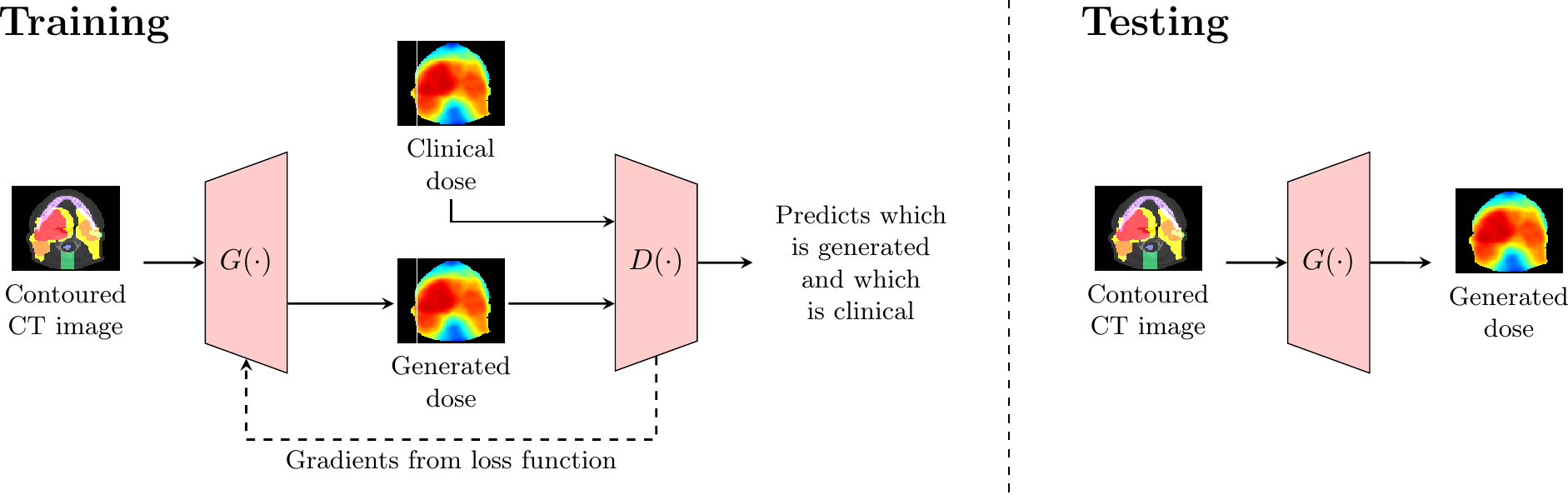} 
        \caption{Overview of the GAN training and testing phases.}
          \label{fig:gan_training}
\end{figure}

The generator and discriminator networks were trained iteratively using gradient descent. After training was complete, the discriminator was disconnected and the generator was used on out-of-sample CT images. Figure~\ref{fig:gan_training} summarizes the difference between how the GANs were trained and tested. In our experiments, we used the loss function given by $\mathcal{L}(D,G)$ with $\lambda=90$, and trained the networks using the Adam optimizer~\citep{kingma:2014adam} with learning rate $\alpha=0.0002$ and $\beta_1=0.5$ and $\beta_2=0.999$. These hyperparameters are the default Adam settings and have been used in many style transfer problems~\citep{isola:2017image}. We also performed a parameter sweep for $\lambda$ and found the default setting sufficient. We stopped training when the standard GAN loss function $\EX_{\pdata}\left[ \log D(\bx, \bc) \right] + \EX_{\pz}\left[ \log (1-D(G(\bz, \bc), \bc)) \right]$ was roughly equal to the regularization term $\lambda \EX_{\pdata,\pz} \left[ \| \bx - G(\bz, \bc) \|_1 \right]$; if the loss from the $l_1$ penalty fell too low, the GAN began to memorize the dataset. Therefore, the 2D-dose and 2D-RGB GANs were trained for $50$ epochs and the 3D-dose GAN was trained for $200$ epochs. It is natural for the 3D-dose GAN to require more epochs to converge as it contains significantly more parameters and generates the entire dose distribution. The code for all experiments is provided at~\url{http://github.com/rafidrm/gancer}.

\subsection{Training the GAN}
We obtained plans for 217 oropharyngeal cancer treatments delivered at a single institution with 6 MV, step-and-shoot, intensity-modulated radiation therapy. Plans were prescribed 70 Gy and 56 Gy in 35 fractions to the gross disease (PTV70) and elective target volumes (PTV56), respectively; in 130 plans there was also a prescription of 63 Gy to the intermediate risk target volume (PTV63). The organs-at-risk (OARs) included the brainstem, spinal cord, right parotid, left parotid, larynx, esophagus, mandible, and the limPostNeck, which is an artificial structure for sparing the posterior neck. Patient geometry was discretized into voxels of size $4$ mm $\times$ $4$ mm $\times$ $2$ mm. 

The CT images and dose distributions for all 217 treatment plans were converted into a suitable format for use by the GAN architectures. The CT images were reconstructed so that each voxel had RGB channels, which were assigned values according to Table \ref{table:Colors}, and converted into 128 axial slices of $128 \times 128$ voxels. The images were separated into a training set of 130 random samples (a total of 15,657 pairs of CT image slices and dose distributions) and a testing set of the remaining 87 samples for out-of-sample evaluation. 

\begin{table}[tb]
\centering
\caption{Colors assigned to each voxel. Voxels that were classified as both OAR and target were assigned nonzero green and red channel values, respectively.}
	\begin{tabular}{l c c c}
	  Structure & Red Channel & Green Channel & Blue Channel \\ \hhline{====}
	Brainstem & 0 & 125 &  CT Grayscale \\	
	Spinal Cord	& 0 & 147 &  CT Grayscale \\	
   	Right Parotid & 0 & 190 &  CT Grayscale \\	
	Left Parotid & 0 & 190 &  CT Grayscale \\	
 	Larynx &  0 & 233&  CT Grayscale \\	
  	Esophagus  & 0 & 212 &  CT Grayscale \\
	Mandible & 0 & 168 &  CT Grayscale \\	
	limPostNeck & 0 & 255 &  CT Grayscale \\	
    	PTV70 & 255 & 0 &  CT Grayscale \\	
  	PTV63 & 205 & 0 &  CT Grayscale \\	
 	PTV56 & 155 & 0 &  CT Grayscale \\	
	Unclassified & 0 & 0 &  CT Grayscale \\	
 	Empty Space & 0 & 0 &  0 \\	
	\end{tabular}
\label{table:Colors}
\end{table}

\subsection{Creating Plans Using Inverse Planning}
During out-of-sample testing, predictions produced by the generator were used as input into an optimization model with two stages. In the first stage, given a predicted dose distribution, the objective weights for a standard inverse planning model were estimated using a parameter estimation technique that has been previously validated in oropharynx~\citep{Babier:2018a}.
In the second stage, the estimated weights were used in an inverse planning optimization model to generate treatment plans. The objective minimized the sum of 65 functions: seven per OAR and three per target. In our experiments, objectives for the OARs included the mean dose, max dose, and the average dose above 0.25, 0.50, 0.75, 0.90, and 0.975 of the maximum predicted dose to the OAR. Objectives for the target included the maximum dose, average dose below prescription, and average dose above prescription. The complexity of all generated treatment plans was constrained to a sum-of-positive-gradients (SPG) value of 55~\citep{Craft:2007aa}. SPG was used since it is a convex surrogate for the physical deliverability of a plan and the parameter 55 was chosen as it is two standard deviations above the average SPG~\citep{Babier:2018b}. The dose influence matrices required for the optimization model were derived with \texttt{A\ Computational\ Environment\ for\ Radiotherapy\ Research}~\citep{CERR}. Each of the KBP-generated plans were delivered from nine equidistant coplanar beams at angles 0$^\circ$, 40$^\circ$, \ldots, 320$^\circ$. We used \texttt{Gurobi 7.5} to solve the optimization model.

We also generated plans using a scaling procedure that multiplicatively increased  the entire predicted dose distribution by the smallest amount to satisfy all target dose criteria. The scaled predictions were then input into the optimization model. Note that this scaling does not affect the fairness of the analysis because the final KBAP plans must satisfy the same constraints (e.g., SPG) as plans generated from the unscaled predictions. To study the full effect of the scaling step, we generated three additional populations of \emph{scaled} final plans alongside the initial three \emph{unscaled} plans corresponding to 2D-RGB, 2D-dose, and 3D-dose. We refer to the three scaled KBAP plans as 2D-RGB$'$, 2D-dose$'$, and 3D-dose$'$ respectively.

\subsection{Performance Analysis}
We conducted two primary analyses. First, we evaluated the quality of the KBAP plans by computing the fraction of clinical planning criteria that were satisfied. We compared these results against the performance of the clinical plans. Second, we evaluated the quality of the KBP predictions by comparing the predicted dose distributions to clinical dose distributions, and the KBAP plan DVHs to their respective predicted DVHs. 

\textbf{KBAP Plan Quality:}
The quality of the final KBAP plans were measured by evaluating how often they satisfied the clinical criteria presented in Table~\ref{goals}. For each class of clinical criteria, i.e., OARs, targets, and all regions-of-interest (ROIs), which includes both OARs and targets, we generated confusion matrices to compare the KBAP plans with the clinical plans. We then analyzed the organ-specific criteria that was achieved by each clinical plan to determine whether the corresponding KBAP plan also passed that criterion. 

\begin{table}[tb]
\caption{The planning criteria used for evaluation: $\mathcal{D}_{99}$ is the dose to a fractional volume of $0.99$, $\mathcal{D}_{mean}$ is the mean dose to a structure, and $\mathcal{D}_{max}$ is the max dose to a structure.}
\centering
	\begin{tabular}{c c}
	  Structure & Criteria \\ \hhline{==}
	Brainstem & $\mathcal{D}_{max} \le$ 54 Gy \\
	Spinal Cord & $\mathcal{D}_{max} \le$ 48 Gy \\
   	Right Parotid & $\mathcal{D}_{mean}\le$ 26 Gy \\
	Left Parotid & $\mathcal{D}_{mean}\le$ 26 Gy \\
 	Larynx & $\mathcal{D}_{mean} \le$ 45 Gy \\
  	Esophagus & $\mathcal{D}_{mean} \le$ 45 Gy \\
	Mandible & $\mathcal{D}_{max}\le$ 73.5 Gy \\
    	PTV70 & $\mathcal{D}_{99}\ \ge$ 66.5 Gy \\
  	PTV63 & $\mathcal{D}_{99}\ \ge$ 59.9 Gy \\
 	PTV56 & $\mathcal{D}_{99}\ \ge$ 53.2 Gy
	\end{tabular}
\label{goals}
\end{table}

\textbf{KBP Prediction Quality:}
We measured KBP prediction quality by evaluating how similar the KBP predictions were to their corresponding KBAP plans. For every ROI and every patient, we calculated the average absolute error between the KBP predicted DVH and the KBAP plan DVH. Differences between 2D-RGB and 2D-dose, 2D-dose and 2D-dose$'$, and 2D-dose$'$ and 3D-dose$'$ were evaluated; each comparison attempts to isolate the impact of each of the three KBAP-specific modifications. We then evaluated the difference between 2D-RGB and 3D-dose$'$ to quantify the joint contribution of all three modifications. The distribution of the errors over the population of plans was visualized using a separate boxplot for each set of comparisons. Finally, we visualized the predicted dose distributions to determine whether the 3D model produced smoother predictions across the longitudinal axis than a 2D model.

\section{Results}

\subsection*{KBAP Plan Quality}\label{Sec:planQual} 
In Table~\ref{clinCritConfusion}, we present confusion matrices comparing the quality of the final KBAP plans with the clinical plans. The columns represent KBAP performance using each of the six KBP approaches while the rows indicate the clinical plans and the performance targets. Overall, 3D-dose$'$ plans best replicated the performance of the clinical plans since they agreed most on what criteria passed and failed (i.e., Pass/Pass and Fail/Fail). Where they differed, 3D-dose$'$ plans satisfied five times as many criteria (Fail/Pass) as the clinical plans (Pass/Fail). We also observed that scaling made a substantial difference as scaled plans outperformed their unscaled counterparts. For example, scaled 3D plans satisfied 99.5\% of all target criteria whereas only 52.3\% were satisfied for unscaled 3D plans. Finally, 2D-dose$'$ and 3D-dose$'$ performed the best satisfying 77.0\% and 76.4\% of all ROI criteria, respectively.

\setlength{\tabcolsep}{0.35em}	

\begin{table}[tb]																																
\centering																																
\caption{For each KBAP approach, the percentage of clinical criteria that passed and failed compared to the corresponding clinical plans.}																																
\begin{tabular}{c l | c | c c | c c | c c |c c |c c |c c}																					\hhline{~~~============}													
		\multicolumn{3}{c|}{}  & 		\multicolumn{6}{c|}{Unscaled}			&			\multicolumn{6}{c}{Scaled}			\\
\hhline{~~~============}												
			\multicolumn{3}{c|}{} 	&	 \multicolumn{2}{c|}{2D-RGB} 			&	\multicolumn{2}{c|}{2D-dose} 			&	 \multicolumn{2}{c|}{3D-dose} 			&	 \multicolumn{2}{c|}{2D-RGB$'$} 			&	\multicolumn{2}{c|}{2D-dose$'$} 			&	 \multicolumn{2}{c}{3D-dose$'$} 	\\	\cline{4-15}
		\multicolumn{3}{c|}{} 	&	Pass	&	Fail	&	Pass	&	Fail	&	Pass	&	Fail	&	Pass	&	Fail	&	Pass	&	Fail	&	Pass	&	Fail	\\	\cline{1-15}	
\parbox[t]{4mm}{\multirow{6}{*}{\rotatebox[origin=c]{90}{Clinical}}}	&		\multirow{ 2}{*}{OARs}	&	Pass	&	63.4	&	3.2	&	64.9	&	1.7	&	65.8	&	0.8	&	60.5	&	6.1	&	63.1	&	3.5	&	63.4	&	3.2	\\		
	&			&	Fail	&	6.2	&	27.2	&	7.9	&	25.5	&	7.9	&	25.5	&	4.7	&	28.7	&	5.9	&	27.5	&	4.6	&	28.8	\\		\cline{2-15}
	&		\multirow{ 2}{*}{Targets}	&	Pass	&	45.5	&	23.2	&	46.9	&	21.9	&	43.8	&	25.0	&	60.3	&	8.5	&	67.9	&	0.9	&	68.3	&	0.4	\\		
	&			&	Fail	&	9.8	&	21.4	&	9.4	&	21.9	&	8.5	&	22.8	&	21.4	&	9.8	&	30.4	&	0.9	&	31.2	&	0.0	\\ \cline{2-15}
	&		\multirow{ 2}{*}{All ROIs}	&	Pass	&	58.5	&	8.7	&	60.0	&	7.2	&	59.7	&	7.5	&	60.5	&	6.7	&	64.4	&	2.8	&	64.7	&	2.4	\\		
	&			&	Fail	&	7.2	&	25.6	&	8.3	&	24.5	&	8.1	&	24.7	&	9.3	&	23.5	&	12.6	&	20.2	&	11.9	&	20.9	\\	
\cline{1-15}
\label{clinCritConfusion}																																
 \end{tabular}																																
\end{table}																										
Table \ref{clinCrit} summarizes the performance of the KBAP plans, focusing only on the criteria that the corresponding clinical plans also passed. Both 3D-dose and 3D-dose$'$ performed the best across all OAR and target criteria. In particular, 3D-dose$'$ achieved the highest passing rate for all target criteria. It also achieved the highest passing rate for all OAR criteria except the larynx and mandible. However, for some regions such as the brainstem, esophagus, and PTV63, multiple KBAP approaches yielded the top result.
															
\setlength{\tabcolsep}{0.5em}	
\begin{table}[tb]													
\centering													
\caption{The percentage of criteria satisfied in each KBAP plan conditional on the clinical plan also achieving them. For each criteria, the highest value in each row is bolded.}													
\begin{tabular}{c l | c c c | c c c c }	
\hhline{~~=======}													
	&	&		\multicolumn{3}{c|}{Unscaled}			&			\multicolumn{3}{c}{Scaled}			\\
\hhline{~~======}													
&	&	2D-RGB	&	2D-dose	&	3D-dose	&	2D-RGB$'$	&	2D-dose$'$	&	3D-dose$'$	\\
\hline													
\parbox[t]{4mm}{\multirow{7}{*}{\rotatebox[origin=c]{90}{OARs}}} & Brainstem	&	\textbf{100.0}	&	\textbf{100.0}	&	\textbf{100.0}	&	\textbf{100.0}	&	\textbf{100.0}	&	\textbf{100.0}	\\
& Spinal Cord	&	\textbf{100.0}	&	\textbf{100.0}	&	\textbf{100.0}	&	98.9	&	98.9	&	\textbf{100.0}	\\
& Right Parotid	&	\textbf{94.1}	&	88.2	&	\textbf{94.1}	&	\textbf{94.1}	&	88.2	&	\textbf{94.1}	\\
& Left Parotid	&	63.6	&	\textbf{81.8}	&	\textbf{81.8}	&	54.5	&	\textbf{81.8}	&	\textbf{81.8}	\\
& Larynx	&	91.8	&	89.8	&	\textbf{98.8}	&	87.8	&	87.8	&	91.8	\\
& Esophagus	&	\textbf{100.0}	&	\textbf{100.0}	&	\textbf{100.0}	&	\textbf{100.0}	&	\textbf{100.0}	&	\textbf{100.0}	\\
& Mandible	&	84.8	&	\textbf{98.5}	&	\textbf{98.5}	&	65.2	&	84.8	&	81.8	\\\hline
\parbox[t]{4mm}{\multirow{3}{*}{\rotatebox[origin=c]{90}{Targets}}} & PTV70	&	48.3	&	82.8	&	79.3	&	75.9	&	\textbf{100.0}	&	\textbf{100.0}	\\
& PTV63	&	\textbf{100.0}	&	\textbf{100.0}	&	94.0	&	\textbf{100.0}	&	\textbf{100.0}	&	\textbf{100.0}	\\
& PTV56	&	52.2	&	15.2	&	10.9	&	89.1	&	95.7	&	\textbf{97.8}	\\ \hline
\label{clinCrit}													
 \end{tabular}													
\end{table}														

In Table~\ref{clinCritPlans}, we recorded the proportion of KBAP plans that satisfied all of the same criteria as the corresponding clinical plans. Note that the subtle difference between these results and the results from Table~\ref{clinCrit} is that here we only record a pass if the KBAP plan meets every single criteria that the clinical plan meets. Table~\ref{clinCrit}, on the other hand, considers each criterion independently. We found that plans generated from the scaled predictions performed better than their unscaled counterparts in terms of satisfying all ROI criteria; we observed the biggest improvement between 3D-dose and 3D-dose$'$ (34.5 percentage points). Like all scaled plans, the improvement of 3D-dose$'$, which satisfied the most target criteria (98.9\%), was the result of much better target coverage at the expense of less OAR sparing compared to the 3D-dose plans, which satisfied the most OAR criteria (94.3\%). Dose encoded as a grayscale image led to an increase in plan quality best shown by the difference of 27.6 percentage points separating 2D-RGB$'$ and 2D-dose$'$. Finally, it was clear that the 3D GAN architecture resulted in large improvements in prediction quality as compared to the 2D approach. That is, 3D-dose$'$ satisfied criteria more frequently (by 2.3 percentage points) than 2D-dose$'$. Overall, 3D-dose$'$ achieved the same criteria as the clinical plans in 78.2\% of cases, which was more than any other approach.

\begin{table}[tb]													
\centering													
\caption{The percentage of plans in each KBAP population that satisfied all clinical criteria that were satisfied by the clinical plans. The highest percentage of satisfied criteria is bolded in each row.}		

\newcommand\Bstrut{\rule[-0.9ex]{0pt}{0pt}}   

\begin{tabular}{l l | c c c | c c c  }													
\hhline{~~======}			
&	&			\multicolumn{3}{c|}{Unscaled}			&			\multicolumn{3}{c}{Scaled}			\\
\hhline{~~======}													
&	&	2D-RGB	&	2D-dose	&	3D-dose	&	2D-RGB$'$	&	2D-dose$'$	&	3D-dose$'$	\\
\hline													
&OARs	&	80.5	&	88.5	&	\textbf{94.3}	&	62.1	&	78.2	&	79.3	\\
&Targets	&	51.7	&	52.9	&	47.1	&	78.2	&	97.7	&	\textbf{98.9}	\\
&All ROIs &	42.5	&	47.1	&	43.7	&	48.3	&	75.9	&	\textbf{78.2}	\\ \hline
\label{clinCritPlans}													
 \end{tabular}													
\end{table}

\subsection*{KBP Prediction Quality} 
\begin{figure}[tb]
 \centering
\subfigure[\ 2D-RGB vs 2D-dose plans]{
\includegraphics[width=0.4\linewidth]{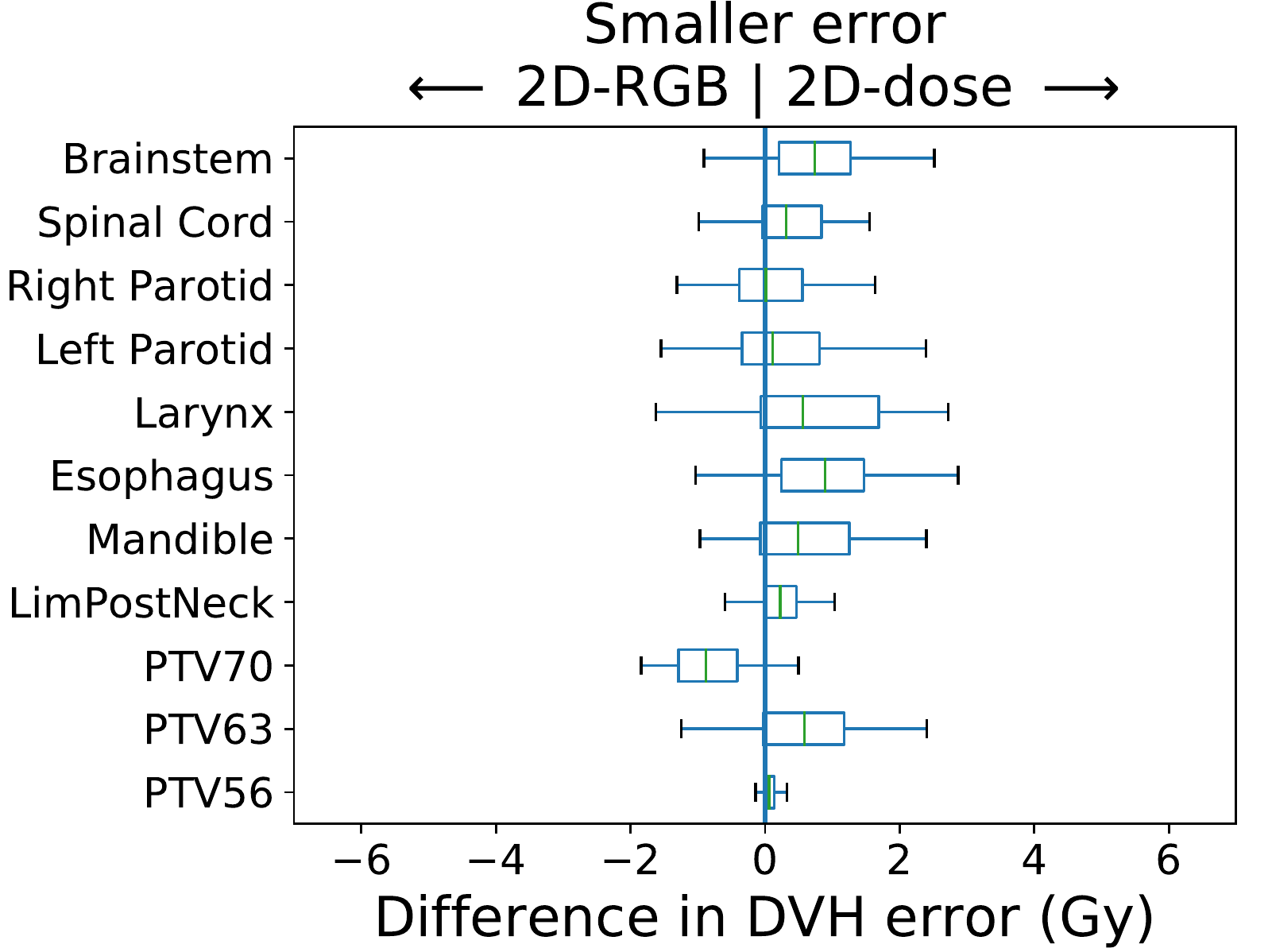} \label{G2RGB}}
\subfigure[\ 2D-dose vs 2D-dose$'$ plans]{
\includegraphics[width=0.4\linewidth]{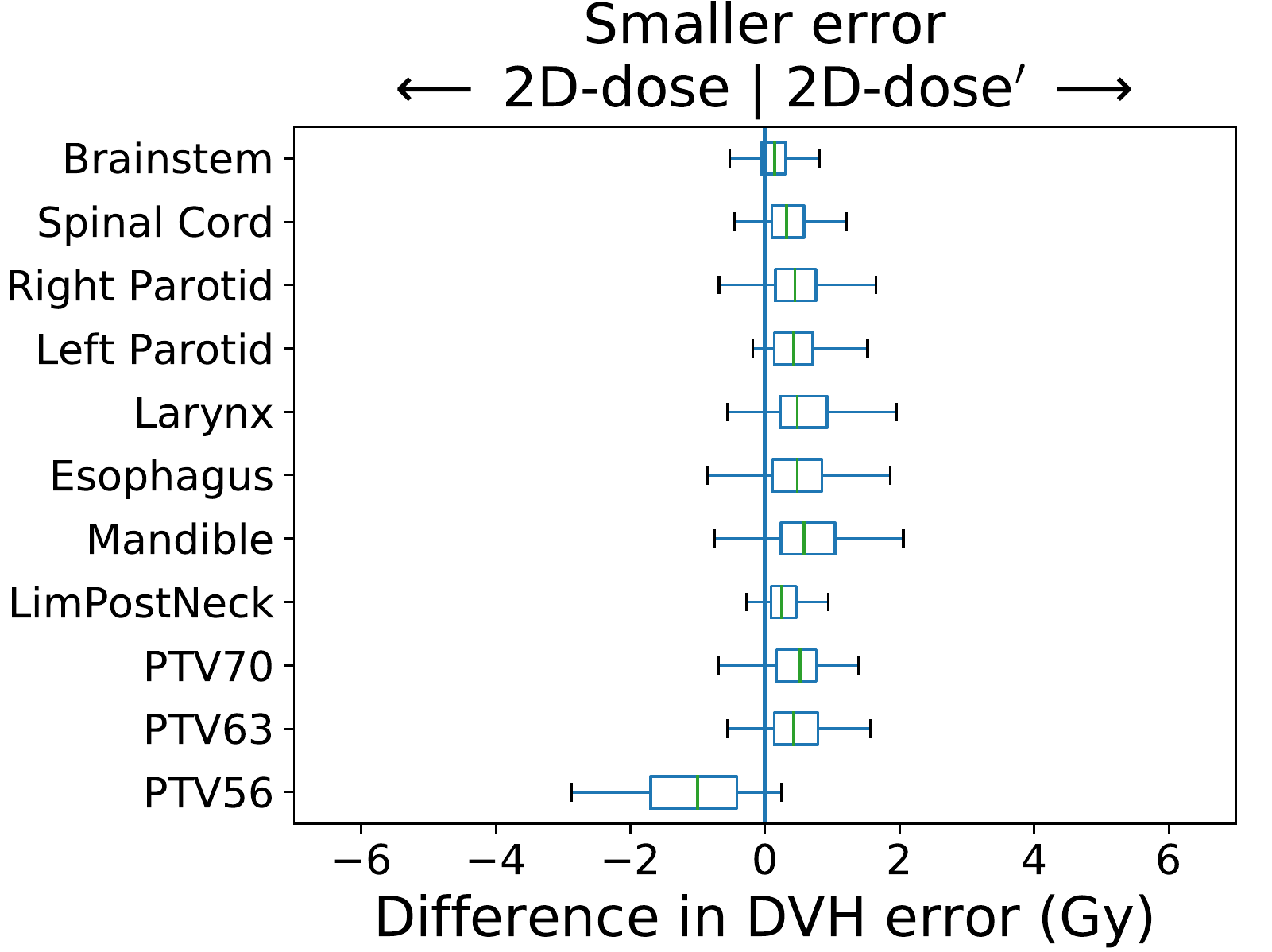} \label{2dScaled}}
\subfigure[\ 2D-dose$'$ vs 3D-dose$'$ plans]{
\includegraphics[width=0.4\linewidth]{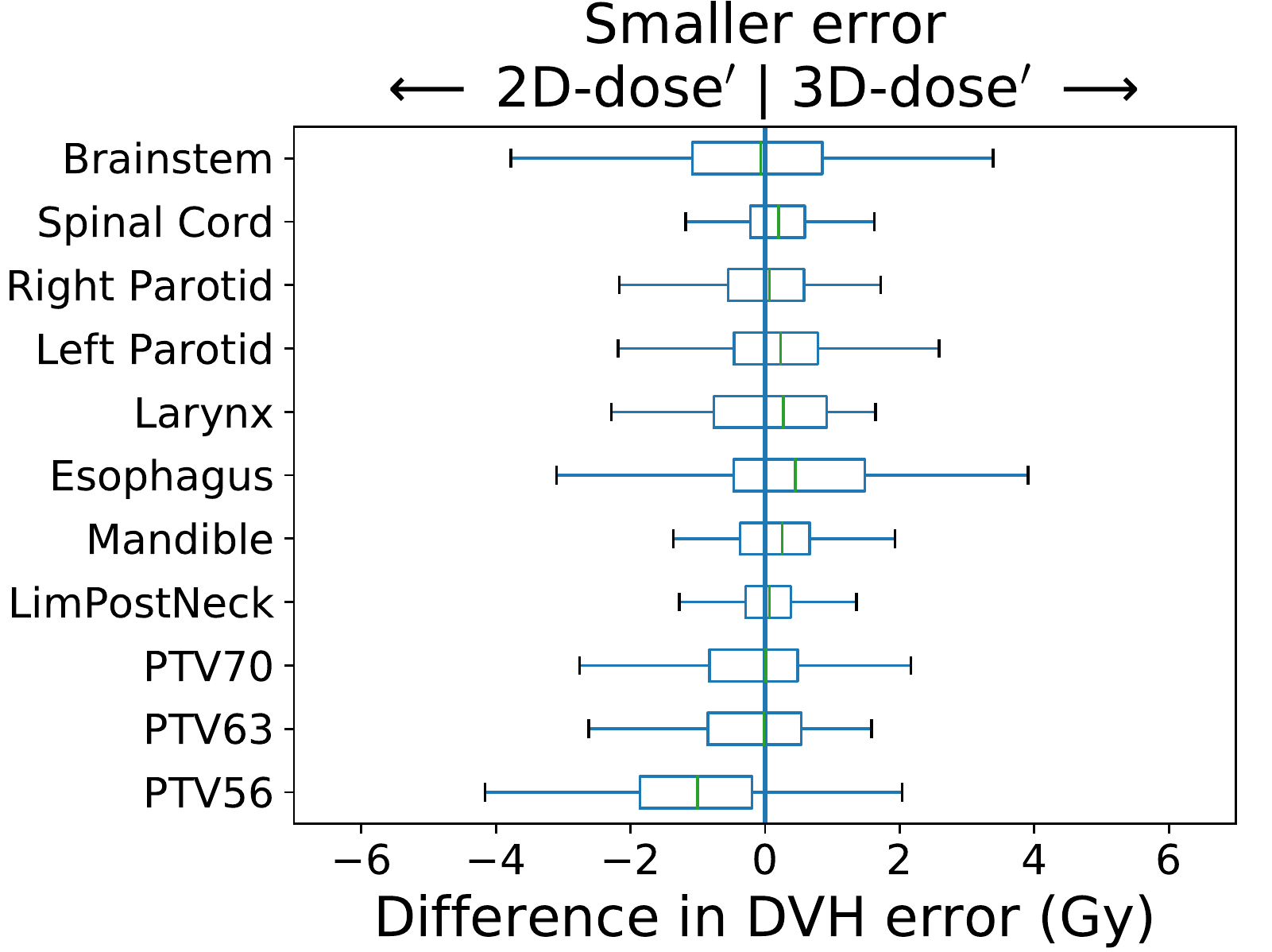}\label{3dScaled} }
 \subfigure[\ 2D-RGB vs 3D-dose$'$ plans]{
\includegraphics[width=0.4\linewidth]{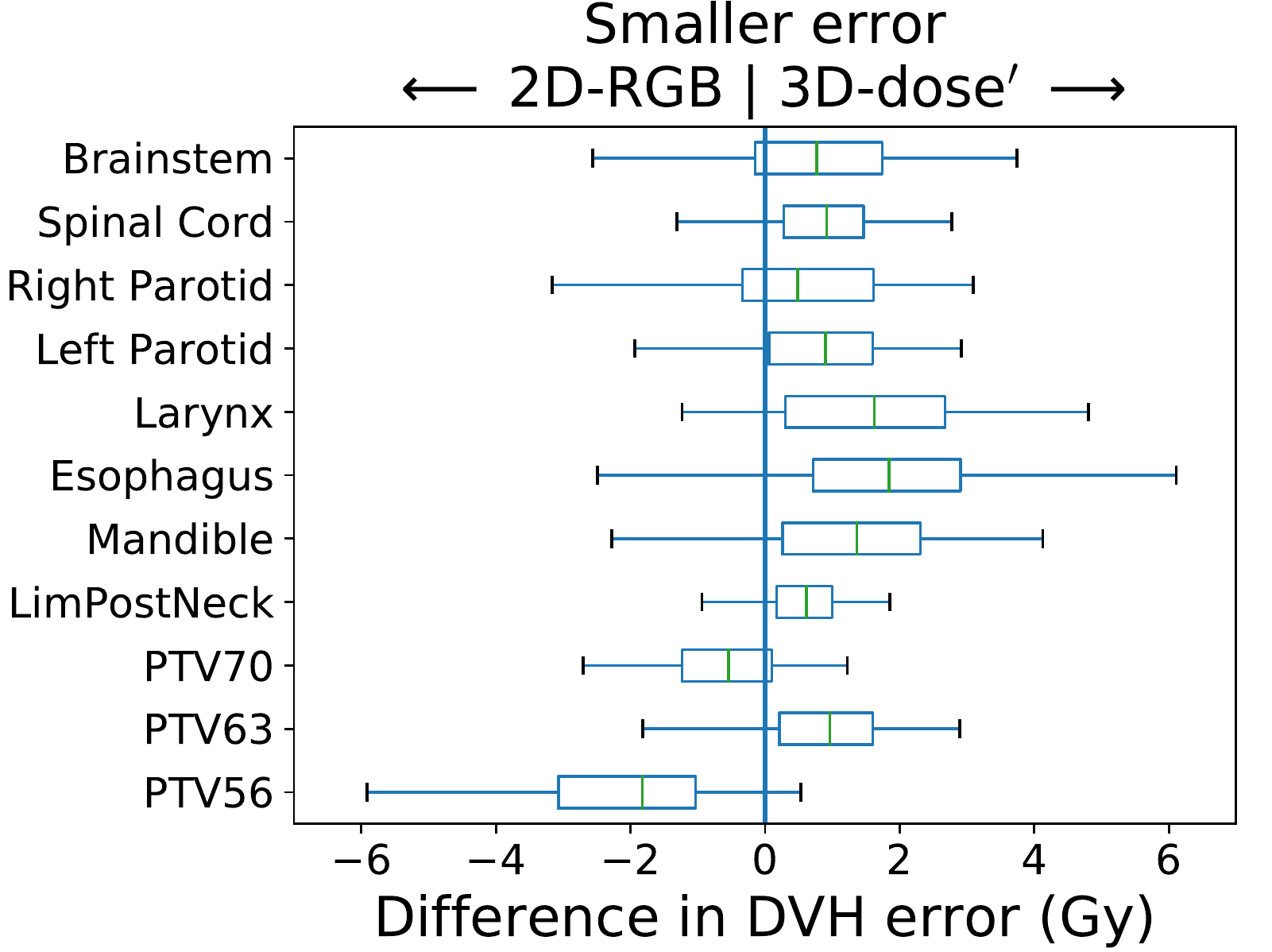} \label{2dV3d}}
\caption{The distribution of DVH errors between KBAP plans with respect to DVHs from their respective KBP predictions. The boxes indicate median and interquartile range (IQR). Whiskers extend to the minimum of 1.5 times the IQR and the most extreme outlier.}
\label{DVHError}
\end{figure}

In Figure~\ref{DVHError}, we present box plots comparing the effect of each modification on prediction error, i.e., 
the average absolute error between the KBP predicted DVH and the KBAP plan DVH. Each subplot represents a pairwise comparison of the prediction error between KBAP pipelines with different GAN architectures for prediction. The larger the skew towards one side, the more accurate that GAN architecture is. By comparing 2D-RGB to 2D-dose, plotted in Figure~\ref{G2RGB}, we found that dose encoded as a scalar value resulted in predictions that were closer (median error difference of 0.19 Gy) to deliverable plans as compared to predictions where dose was encoded as a RGB color map. We also found that the prediction errors were lower among scaled plans than unscaled plans (Figure~\ref{2dScaled}). That is, by comparing 2D-dose and 2D-dose$'$, we found that half of all predictions generated via scaling were about 0.31 Gy more accurate than those generated without; the PTV56 was the only structure where the scaled KBAP plans were less accurate (average median error of 1.00 Gy). For both modifications, although the median differences were modest, the upper limit of the differences could be upwards of 2 Gy. Additionally, we found that the median error of 3D-dose$'$ was 0.07 Gy lower than  2D-dose$'$ (Figure~\ref{3dScaled}). Finally, in Figure~\ref{2dV3d}, we evaluated the difference between our best set of KBAP plans, 3D-dose$'$, and the plans generated by the previous state-of-the-art in GAN-based KBAP, 2D-RGB. The median error of 3D-dose$'$ was 0.63 Gy smaller than 2D-RGB, which suggests that the predictions made by 3D-dose$'$ were closer to the final dose distributions than those generated by 2D-RGB.

Figure~\ref{doseWashes} presents predicted dose distributions corresponding to the 2D-dose and a 3D-dose KBP methods. Note that the two models predicted distinct dose distributions that differed along the vertical (i.e., longitudinal) axis. In particular, the 3D GAN better learned the vertical relationship between adjacent axial slices and thus, generated smoother dose predictions across the longitudinal axis. In contrast, the 2D GAN predictions included more ``streaky'' and unrealistic discontinuities in the dose distribution, particularly around axial slices that were adjacent to the target boundaries. For example, the dose falloff was impossibly steep in the plans generated using 2D predictions (Figure~\ref{2D-doseWash}), while plans generated using 3D predictions had more realistic dose gradients (Figure~\ref{3D-doseWash}).

\section{Discussion}

\begin{figure}[t!]
\centering
\subfigure[\ CT image]{
\includegraphics[width=0.4\linewidth]{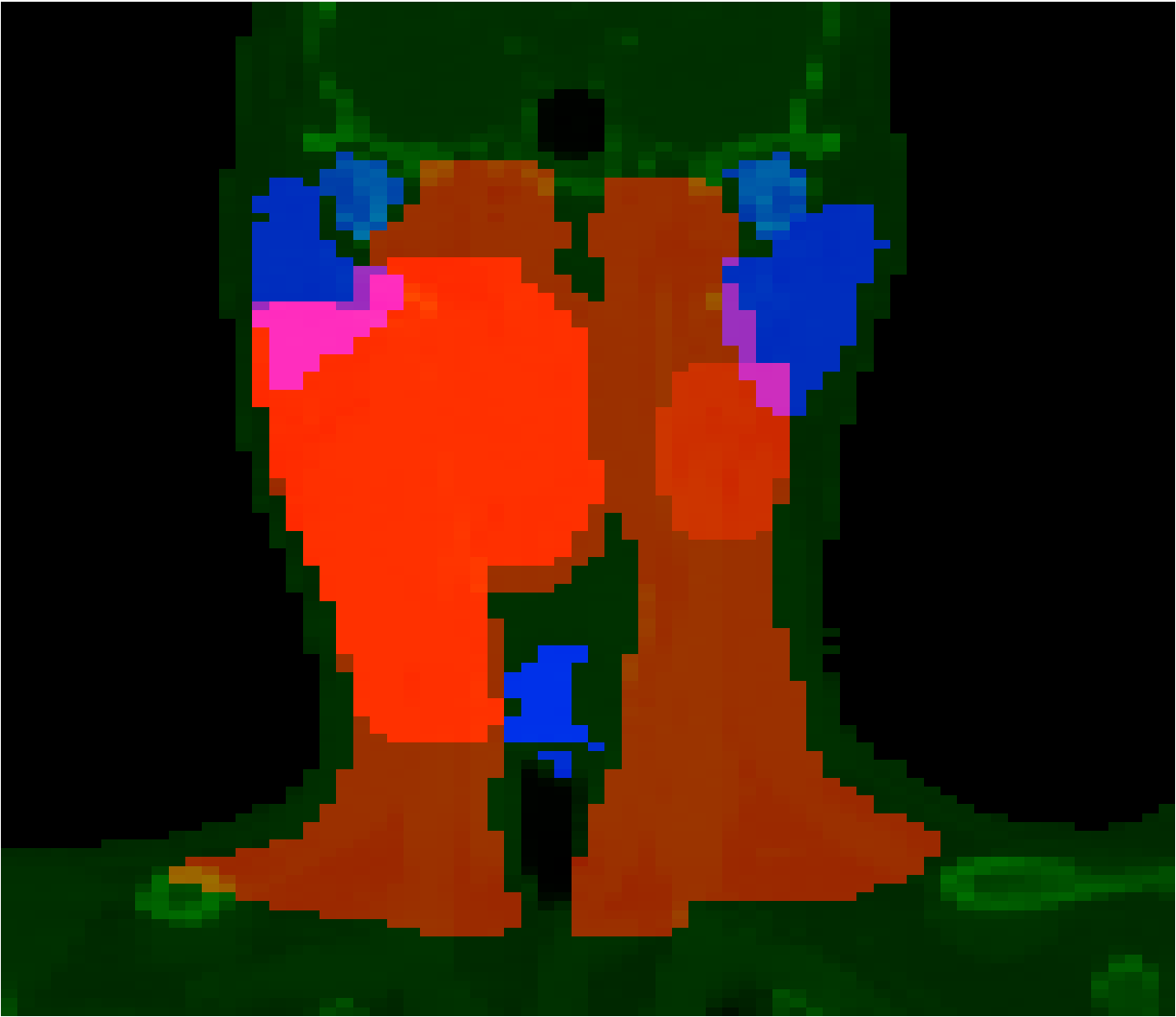} }
\subfigure[\ Clinical plan]{
\includegraphics[width=0.4\linewidth]{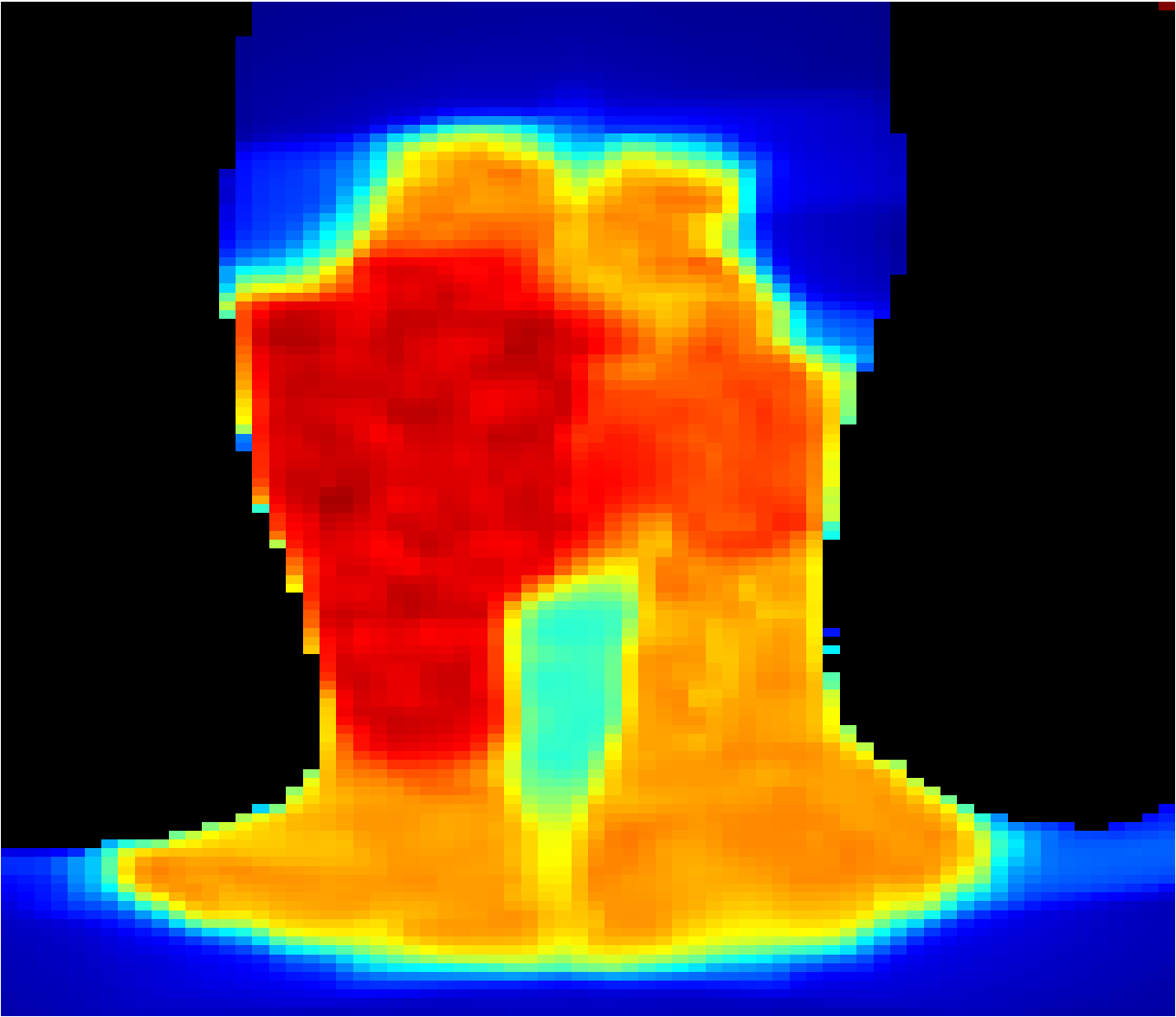}}
\subfigure[\ 2D-dose prediction]{
\includegraphics[width=0.4\linewidth]{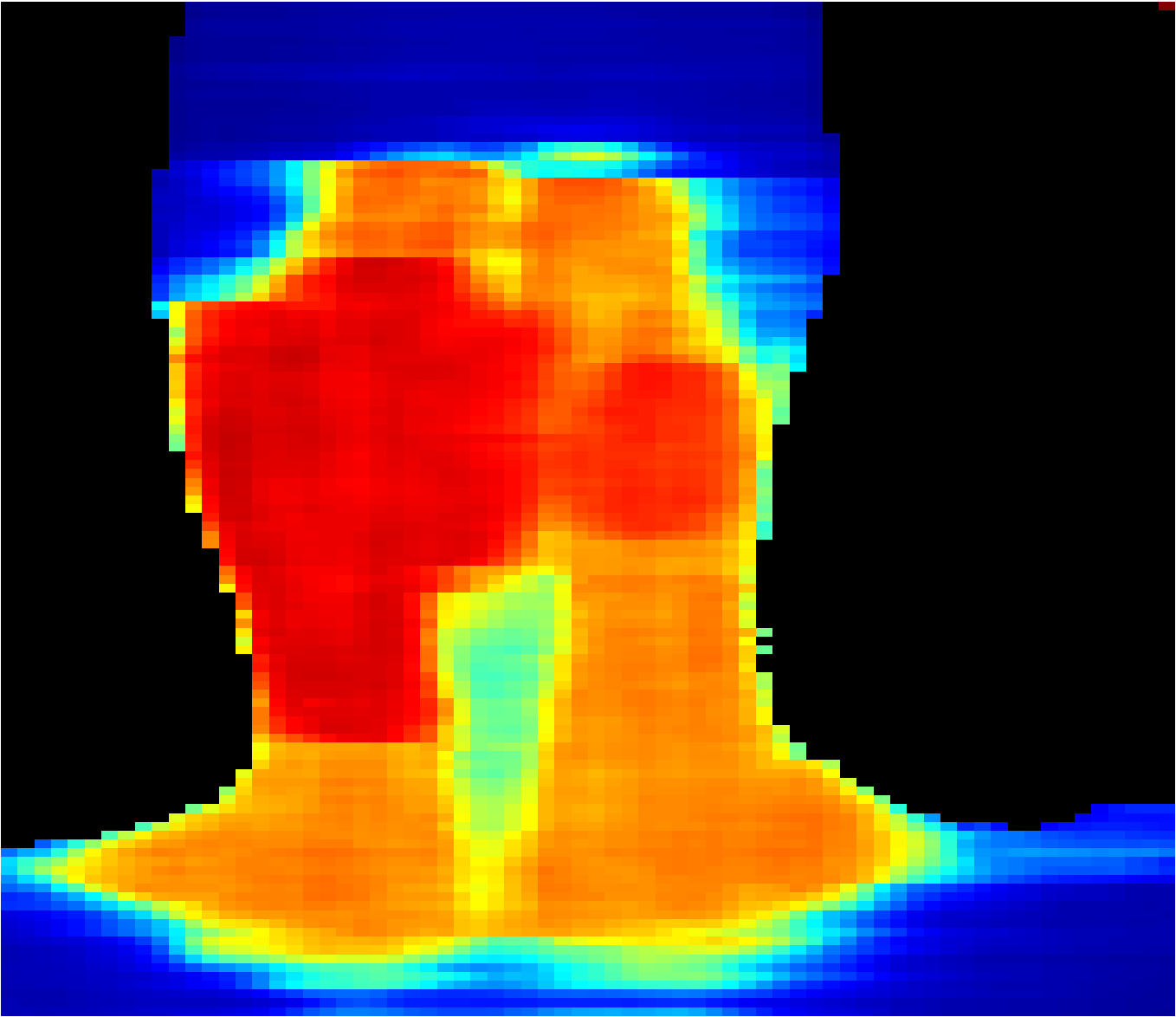} \label{2D-doseWash}  }
\subfigure[\ 3D-dose prediction]{
\includegraphics[width=0.4\linewidth]{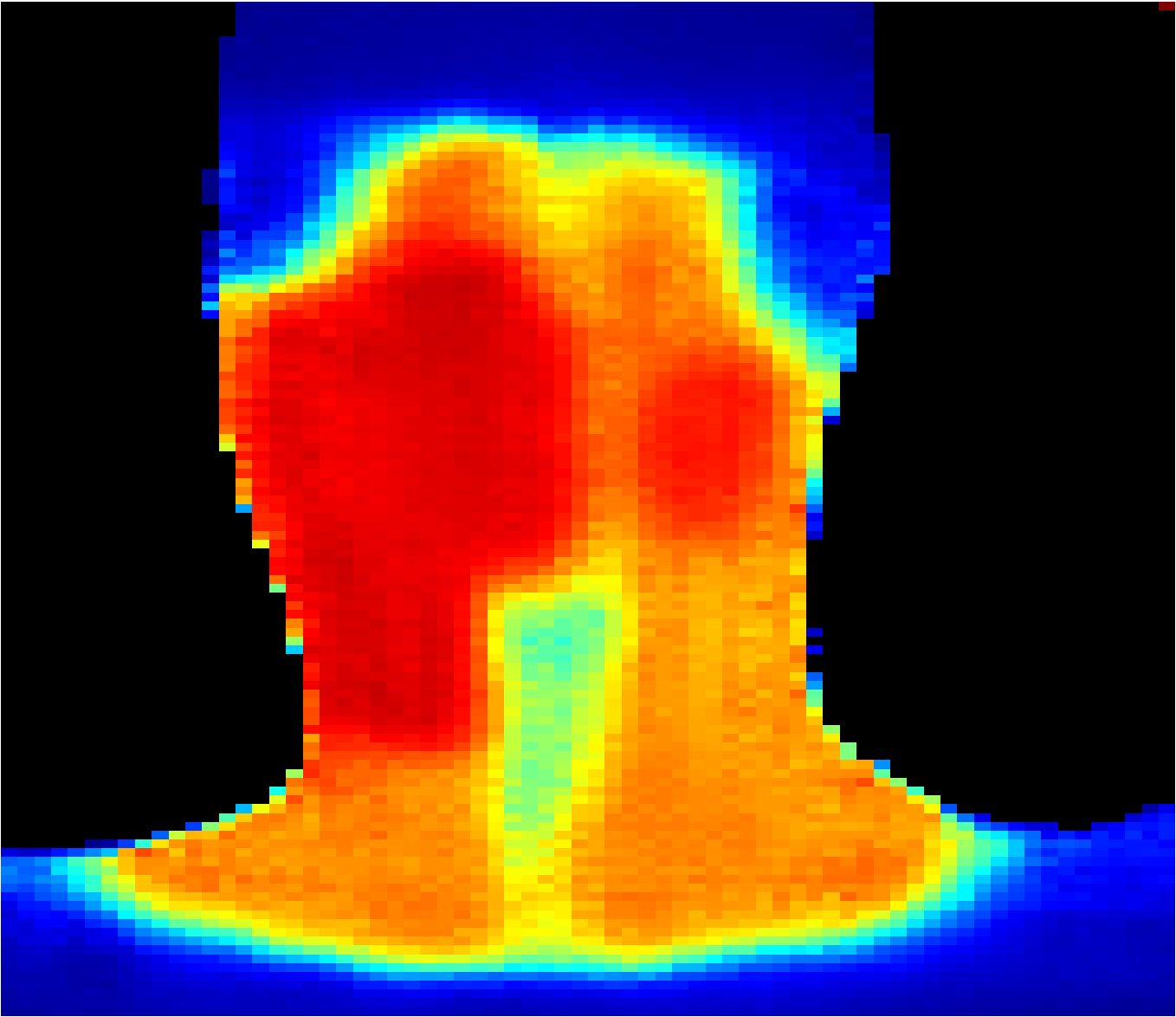} \label{3D-doseWash} }
\includegraphics[trim={0 0 0 11.5cm},clip, width = 0.45\linewidth]{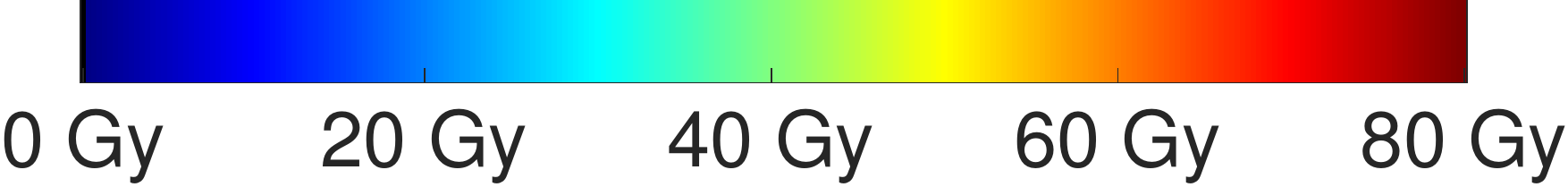}
\caption{The dose distributions for a sample patient over a single CT image (a) of their clinical plan (b), 2D-dose prediction (c), and 3D-dose prediction (d).}
\label{doseWashes}
\end{figure}

Although historically, KBP methods have predicted DVHs using hand-tailored features, there is widespread interest in automatically predicting dose distributions~\citep{McIntosh:2016aa,nguyen2017dose,GANCER,Fan:2018aa}. In this paper, we extend the literature by building a KBAP pipeline that automatically generates treatment plans from CT images. The pipeline consists of two major components: prediction and optimization. The prediction stage uses a generative adversarial network to predict dose distributions from a CT image. The optimization stage consists of two optimization models, a parameter estimation model that learns objective function weights from the predicted dose distribution and an inverse planning model that produces the final fluence-based treatment plans. We demonstrate that our approach produces treatment plans that are better than the previous state-of-the-art in KBAP and generates full dose distributions rather than summary statistics~\citep{McIntosh:2017ab,Fan:2018aa}.

Our framework includes three enhancements that improved performance: 1) representing dose to each voxel as a scalar value, 2) scaling the KBP prediction from the GAN prior to optimization, and 3) predicting the full 3D dose distribution from 3D CT images.

\textbf{Representing dose to each voxel as a scalar.} We hypothesized that predicting dose encoded in single value (i.e., grayscale) would be easier than predicting dose encoded as a 3-channel RGB value. Our experiments confirmed this result. In particular, 2D-dose$'$ plans satisfied the same criteria as the clinical plans 74\% more often across all ROIs as compared to 2D-RGB$'$ plans. The improvement due to this modification is likely due to the fact that it is much easier to predict a single value rather than a triplet.

\textbf{Scaling the predictions before optimization.} Scaling significantly enhanced the final KBAP plan quality. Scaled KBAP plans satisfied the same criteria as clinical plans 52\% more often than unscaled plans; scaled plans also satisfied 10\% more criteria than the unscaled plans overall. The idea of scaling predictions prior to optimization is novel in the KBAP literature and is a general tool that can be applied to any other KBP method. There are two points worth highlighting. First, scaling is done automatically, just like how our approach automates high-dimensional feature selection. Thus, our KBAP pipeline remains automated. Second, we believe that scaling works because it corrects small inaccuracies that may arise when learning the absolute dose delivered. That is, the GAN seems to be more effective at learning how dose varies among different tissues rather than learning the exact dose that should be delivered to a tissue (otherwise, scaling would not make any difference). 

\textbf{Predicting the full 3D dose distribution from 3D CT images.} Whereas previous work predicted dose to each voxel or axial slice independently and then stitched the predictions together to form the full 3D dose distribution~\citep{Shiraishi:2016ab, nguyen2017dose,GANCER, Fan:2018aa}, our 3D GAN was designed to take an entire contoured 3D CT image as input and generate the corresponding 3D dose distribution as output. Because of this enhancement, the 3D GANs better learned the vertical relationship between adjacent axial slices. Indeed, we observed that the 3D GAN predicted more realistic dose distributions than the 2D GAN (e.g., Figure~\ref{doseWashes}) with smoother dose predictions across the longitudinal axis. 

In our experiments, we used clinical criteria as the primary performance measure to evaluate the plan quality of several GAN architectures. Since it is generally impossible to develop plans that simultaneously achieve all clinical criteria--in our dataset of 217 clinically delivered plans, only 68.4\% of criteria were achieved--our primary goal was to achieve as many criteria as possible. Secondarily, we were interested in generating plans that met the same criteria that the original clinical plans achieved; presumably, these represent the criteria that clinicians originally believed to be the most important. We believe that an automated planning method that produces dose distributions that satisfy the same criteria as treatment plans that have already been delivered is more likely to be clinically implemented.

There are several reasons why we believe GANs are a good choice for KBP. First, they have a history of performing well in applications that involve medical images; specifically in the detection of brain lesions~\citep{alex2017generative} and image augmentation for liver lesion classification~\citep{frid:2018gan}. Second, we found that all of our GAN models performed well inside a KBAP pipeline without significant parameter tuning and architecture modification, both of which are essential and potentially time consuming steps in conventional GAN implementations. We attribute this success to the application; the prediction of dose distributions is akin to the prediction of relatively smooth and uniform images with the same orientation. Third, in the KBAP pipeline, the GAN produces images that are used as input for an optimization model to obtain treatment plans via a traditional inverse planning procedure. Thus, the GAN learns a simpler style mapping as compared to conventional applications, and the optimization phase acts as a safety net to correct potential errors. Finally, it is interesting to note that the method to train a GAN conceptually mimics the iterative process between a treatment planner and an oncologist. The generator behaves as a treatment planner by proposing deliverable dose distributions while the discriminator behaves as an oncologist by determining whether the proposed dose distribution is suitable. 

While other pipelines that predict dose directly have used voxel-based dose mimicking to construct the final plans~\citep{McIntosh:2017aa}, we chose to do inverse planning using DVH-based objectives following prediction because it is in line with current clinical practice and is the most common approach used in the academic KBAP literature~\citep{ Wu:2017aa,Babier:2018b,Fan:2018aa}. We also emphasize that our method does not use hand-tailored feature engineering (e.g., features derived from overlap-volume histograms). 
Thus, as compared to existing KBAP methods, we expect our pipeline to be easier to implement in practice and can result in more predictable results if custom treatment plans are desired. For example, institutions with specific clinical guidelines can train a GAN on images they deem indicative of an ideal treatment plan. In addition, in the future it may be possible that several medical centers combine data to form a large training set, which should further improve performance.

A limitation of our approach is that the prediction and optimization steps are separate stages in the overall pipeline. In theory, an integrated model that does both prediction and treatment plan optimization simultaneously should produce even better results. A second limitation is that our approach requires a clean, well-structured and high-quality dataset, where all images need to have a consistent size (in terms of number of pixels), coloring convention, and orientation. Finally, as with any neural network-based approach, GAN predictions suffer from a lack of interpretability. It is not straightforward to understand why the GAN makes certain predictions, effectively rendering it a black box. Consequently, a treatment planner may have more difficulty using this approach to understand when and why prediction errors occur.

\section{Conclusion}
We developed the first knowledge-based automated planning framework using a 3D generative adversarial network for prediction. Our results based on 217 oropharyngeal cancer treatment plans demonstrated superior performance in satisfying clinical criteria and generated more realistic predictions compared to the previous state-of-the-art. Our three unique contributions to the KBAP architecture of one-channel prediction, post-prediction scaling, and direct prediction of the full 3D dose allowed us to generate high-quality treatments without manual intervention.

\section{Acknowledgments}
This research was supported in part by the Natural Sciences and Engineering Research Council of Canada.
%
\bibliography{refsforMed}
\end{document}